\documentclass[11pt]{article}
\newcommand{\version}{May 9, 2005}                                                  \usepackage{amsthm,amsfonts,amsmath}
\swapnumbers                                
\theoremstyle{plain}                                                                                                     
\newtheorem{thm}{THEOREM}[section]
\newtheorem{lm}[thm]{LEMMA}

\theoremstyle{definition}

\theoremstyle{remark}
\newcommand{\upchi}{\raise1pt\hbox{$\chi$}}
\newcommand{\R}{{\mathord{\mathbb R}}}

\numberwithin{equation}{section}
\begin{document}

\markboth{\scriptsize{CCELM \version}}{\scriptsize{CCELM \version}}

\title{\bf{Droplet minimizers for the Cahn--Hilliard free energy functional}}
\author{\vspace{5pt} E. A.  Carlen$^1$, M. C. Carvalho$^2$, R. Esposito$^3$, 
J. L. Lebowitz$^4$ and R. Marra$^5$\\}

\date{\version}
\maketitle

\centerline{I.H.\'E.S.}

\centerline{Le Bois--Marie, 35 route de Chartres}

\centerline{F-91440 Bures--sur--Yvette, France}

\def\O{\Omega}

\footnotetext                                                                         
[1]{ School of Mathematics, Georgia Tech,
Atlanta, GA 30332, U.S.A. Work partially
supported by U.S. National Science Foundation
grant DMS 03-00349.  }
\footnotetext 
[2]{Department of Mathematics and CMAF, University of Lisbon,
1649-003 Lisbon, Portugal}
\footnotetext 
[3]{Dip. di
Matematica, Universit\`a di L'Aquila, Coppito, 67100 AQ, Italy}
\footnotetext 
[4]{Departments of Mathematics and
Physics, Rutgers University, New Brunswick, NJ
08903, U.S.A.}
\footnotetext 
[5]{Dipartimento di Fisica and Unit\`a INFN, Universit\`a di
Roma Tor Vergata, 00133 Roma, Italy. \\
\copyright\, 2005 by the authors. This paper may be reproduced, in its
entirety, for non-commercial purposes.}
                                                                      
\begin{abstract}

We prove theorems characterizing the minimizers in a model for condensation based on the
Cahn--Hilliard free energy functional. In particular we exactly determine the critical density for droplet
formation.
\end{abstract}

\bigskip
\centerline{Mathematics Subject Classification Numbers:  49S05, 52A40, 82B26}

\section{Introduction} \label{intro}

\subsection{The variational problem} \label{intro1}

Let $\Omega$ be the $d$--dimensional square torus with volume $L^d$.
Consider the free energy functional ${\cal F}(m)$ defined by
\begin{equation}\label{(1)}
{\cal F}(m) = {\theta^2\over 2}\int_\Omega |\nabla m|^2 {\rm d}x +
\int_\Omega F(m){\rm d}x\ ,
\end{equation}
where $\theta$ is a parameter with the units of distance, and 
$$F(m) = {1\over 4}(m^2 -1)^2\ .$$
This is a ``double well'' potential with minima at $m = \pm 1$.   The two minima are the two ``phases''
of the system.    The function $m(x)$ is an ``order parameter field'', representing a summary of the microscopic 
information about the underlying system locally at the point $x$ that is necessary to compute a density for the
Helmholz free energy. The particular free energy density considered here is phenomenological; it does not
arise from any particular microscopic model, but it is a simple caricature of the Helmholz free energy densities  
that do arise from scaling limits of actual microscopic systems with phase transitions. See \cite{CCELMa}
for discussion on the relation with the microscopic models. The free energy functional
(\ref{(1)}) is frequently called the Cahn--Hilliard free energy functional, but it has a fairly ancient history. 
It was already discussed by van der Walls \cite{VW} in the nineteenth
century.

For any number $n$ with $-1 < n < 1$, define the {\em minimal free energy function}  $f_L(n)$ by
\begin{equation}\label{mainmin}
f_L(n) = \inf\left\{\ {\cal F}(m) \ :\ {1\over L^d}\int_\Omega m(x){\rm d}x =n\ \right\}\ .
\end{equation}

In what follows we shall work in units with 
$$\theta =1\ .$$ 
This makes $L$ and $x$ dimensionless. That is, we are implicitly introducing dimensionless coordinates $\tilde x = 
x/\theta$ and a dimensionless scale parameter $\tilde L = L/\theta$. However, setting $\theta =1$, we can drop the
tildes, and work directly in dimensionless coordinates.

If we fix a value of $n$ with $-1 < n < 1$, the minimization problem (\ref{mainmin}) simplifies 
as $L$ becomes very large:   In the bulk of $\Omega$ it must be the  case that   $m(x) \approx  \pm 1$ to high accuracy.    
Consider any order parameter field $m(x)$ that takes on only the two  pure phase values
$\pm 1$. Let $V_+$ be the volume of the region in which $m(x) = 1$. While such an order parameter field
would be discontinuous, and would therefore yield an infinite free energy, The quantity $V_+$ is quite
relevant to (\ref{mainmin}) for large $L$.  From the constraint $\int_\Omega m(x){\rm d}x =nL^d$, 
$V_+ -(L^d-V_+) = nL^d$ so that $V_+$ and $n$ are related by
\begin{equation}\label{Aofn}
V_+ ={n+1\over 2} L^d\qquad{\rm and}\qquad  n = -1 + 2{V_+\over L^d}\ .
\end{equation}

Any order parameter field  $m(x)$ that is a minimizer for (\ref{mainmin}) will be
continuous, and in fact, $C^\infty$ as a consequence of the
Euler--Lagrange equation that it must satisfy. Define the interface
$\Gamma$ between the two phases  by 
$$\Gamma = \{\ x\ :\ m(x) = 0\ \}\ .$$
Let $|\Gamma|$ denote ${\cal H}^{d-1}(\Gamma)$, the  $d-1$ dimensional Hausdorff
measure of $\Gamma$.   Because of the gradient terms in the free energy functional, a minimizer $m$
cannot make a sudden transition for $+1$ to $-1$ in crossing $\Gamma$, and so one would expect to pay a price
for making such a transition that is proportional to $|\Gamma|$.  That is, we would expect that
\begin{equation}\label{surften}
f_L(n) \approx S|\Gamma|
\end{equation}
for some proportionality constant $S$, which is called the {\it surface tension}.  For the physics behind this
terminology and this approximation, see \cite{WR}. 

Accepting this for the moment, if the volume of the region in which $m(x) > 0$ is $V$, we would  expect $\Gamma$ to  be a
surface of minimal area bounding a region of volume $V$. The solution of the isoperimetric problem on the torus $\Omega$
depends on $V/L^d$.  Let $\sigma_d$ denote the surface area of the unit sphere in ${\Bbb R}^d$. Since the volume of the
corresponding unit ball in $\R^d$ is $\sigma_d/d$, a volume of size $V$ can be enclosed either by a sphere  of radius
$(V/(\sigma_d/d))^{1/d}$, or in circular cylinder of radius $s$ and length $L$.  In $d=2$, this is
a ``strip'' between two  parallel lines with length $L$ each.   Because we are in the torus, the caps of the cylinder are not part of the
boundary. Thus the surface of such a cylinder is 
$\sigma_{d-1}s^{d-2}L$, while the volume is $(\sigma_{d-1}/d)S^{d-1}L$. Equating these to the surface area and volume of
a sphere of radius $r$, $\sigma_d r^{d-1}$ and $(\sigma_d/d)r^d$ respectively, we see that both are equal,
when $r = r_c$ where 
\begin{equation}\label{critr}
r_c=( (d-1)/d)^{d-2}(\sigma_{d-1}/\sigma_d)L\ .
\end{equation}
 It pays to take advantage of the periodic boundary condition
by reaching out to the boundary only for very large volume: $V$ must exceed $(\sigma_d/d)r_c^d$, which is proportional to $L^d$, in order for this to be advantageous.  For smaller volumes, the solution of the isoperimetric inequality
is still given by a sphere.

In what follows, we shall be concerned with values of $n$ sufficiently close to $-1$, so that, with $V = V_+$, as  given
by (\ref{Aofn}), the minimal bounding surface is  a sphere.

The {\it equimolar radius}, $r_0$,  is defined to be the radius of the sphere whose enclosed volume is
$V_+$, where $V_+$ is given in terms of $n$ by (\ref{Aofn}).   Since $(\sigma_d/d)r_0^d= V_+$,
we have from (\ref{Aofn}) that \begin{equation}\label{equimolar}
r_0= \left({d\over 2\sigma_d}(n+1)\right)^{1/d}L   \quad= \left({V_+\over \sigma_d/d}\right)^{1/d}\ .
\end{equation}
Here we shall only consider values of $n$ for which 
$\sigma_d r_0^{d-1}  < 2L^{d-1}$, or, in other words,
\begin{equation}\label{erbound}
r_0\le r_c
\end{equation}
with $r_c$ given by (\ref{critr}).

\subsection{Two simple trial functions} \label{intro2}

Under the condition (\ref{erbound}) on $n$ and $L$, there are two natural trial functions to consider for the variational 
problem (\ref{mainmin}).

 The first of these is the {\em equimolar droplet  trial function}
\begin{equation}\label{mdrop}
m_{\rm emd}(x)= m_0\left(|x| - r_0(n)\right)\ ,
\end{equation}
where we are representing $\Omega$ as the centered cube in $\R^d$ with side length $L$ and periodic boundary conditions,  
and $m_0(z)$ is some function such that $\lim_{z\to \pm \infty}m_0(z) = \mp 1$, with the transition from $+1$ to $-1$
being made in such a way as to minimize the cost in free energy.  In fact, we require that the limits $\lim_{z\to \pm
\infty}m_0(z) = \mp 1$ are achieved at finite values of $z$ such that (\ref{mdrop}) does indeed define a smooth function
on $\Omega$.

The second of these is the {\em uniform trial function}
\begin{equation}\label{muni}
m_{\rm uni}(x) = n\ ,
\end{equation}
 corresponding to a ``supersaturated'' state 
with the order parameter strictly between the minimizing values.

Accepting the validity of the approximation (\ref{surften}), we have ${\cal F}(m_{\rm emd}) \approx  S\sigma_dr_0^{d-1}$, 
and therefore, from (\ref{equimolar})
\begin{equation}\label{fmdropapprox}
{\cal F}(m_{\rm emd})\ \approx\  S\sigma_d\left({d\over 2\sigma_d}(n+1)\right)^{1-1/d}L^{d-1}
\ =\  S\sigma_d\left({V_+\over \sigma_d/d}\right)^{1-1/d}\ , 
\end{equation}
one easily computes that
\begin{equation}\label{fmuniapprox}
{\cal F}(m_{\rm uni})\ =\ {1\over 4}(n^2 -1)^2 L^d \  = \ 4{V_+^2\over L^d}\left(1 - {V_+\over L^d}\right)^2\ . 
\end{equation}

Which of these trial functions provides a better description of the minimizers in (\ref{mainmin})?  That depends on
$n$, or what is the same, on the ratio $V_+/L^d$.
In fact, there are now two obvious scaling regimes to consider: We can take $L$ to infinity while keeping either  $n$
or $V_+$  constant.  For this reason, we have expressed ${\cal F}(m_{\rm uni})$ and
${\cal F}(m_{\rm emd})$ in terms of both $n$ and $V_+$. 

If one holds $n$ constant, and takes $L\to \infty$, then  $m_{\rm emd}$ does much better than
$m_{\rm uni}$. On the other hand, if one holds $V_+$ constant as $L$ tends to infinity, we see from (\ref{fmuniapprox})
and (\ref{fmdropapprox}) that   $m_{\rm uni}$ does much better than
$m_{\rm emd}$ for large $L$: This suggests that a  droplet of the $+1$ phase will always ``evaporate'' into the surrounding
$-1$ phase if the  ambient volume $|\Omega|$ is sufficiently large compared to $V_+$.

\subsection{The critical scaling regime} \label{intro3}

The situtation is much more interesting if one considers $f_L(n)$ with $n$ tending towards $-1$ at the same
time that  $L$ tends to infinity:
we seek the smallest value of $n(L)$ for which droplets are stable in a box of volume $L^d$, and seek also to  determine
the structure of such critical minimizing droplets.

In this sort of scaling regime,  $V_+/L^d$ will be very small, and we
can express ${\cal F}(m_{\rm uni})$ in  more physically meaningful terms  as follows: Define
the {\em compressibility} $\chi$ by 
\begin{equation}\label{chidef}
\chi = {1\over F''(-1)}\ .
\end{equation}
 Then since
$F(-1) = F'(-1) = 0$,
$$F(n) = F(-1 + V_+/L^d) \approx {1\over 2\chi}\left({V_+\over L^d}\right)^2\ ,$$
this gives us the approximation
\begin{equation}\label{uniinchi}
{\cal F}(m_{\rm uni}) \approx    {1\over 2\chi}{V_+^2\over L^d}\ .
\end{equation}
Of course, in our problem, $\chi = 1/2$. But introducing the compressibility highlights a competition
between surface and bulk terms in minimizing the free energy.

When $V_+^{1+1/d} \asymp L^d$, 
$$
{\cal F}(m_{\rm emd}) \approx 
S\sigma_d\left({V_+\over \sigma_d/d}\right)^{1-1/d}\qquad{\rm and}\qquad 
{\cal F}(m_{\rm uni}) \approx    {1\over 2\chi}{V_+^2\over L^d}$$
are comparable.   For this reason, we refer to $V_+^{1+1/d} \asymp L^d$ as the {\em critical scaling regime}.
In terms of $n$ and the equimolar radius $r_0$, the critical scaling regime is characterized  by
\begin{equation}\label{critreg}
n+1 \asymp L^{-d/(d+1)}\qquad{\rm or,\ equivalently}\qquad r_0 \asymp L^{d/(d+1)}\ .
\end{equation}
What should one expect for the minimizing free energy  in the critical scaling regime, and will the minimizers
be given by some sort of droplet, or not?

In a recent and incisive  investigation of droplet formation in  $2$ dimensional Ising model \cite{BCK},
Biskup, Chayes and Kotecky  proposed that to answer this question, 
 one should introduce a {\em volume fraction} $\eta$, and put $\eta V_+$ into the drop, and $(1-\eta)V_+$  into the
uniform background.  They then constructed a phenomenological thermodynamic free energy function $\Phi(\eta)$
 which is the sum of the surface tension term and the uniform background term:
\begin{equation}\label{fbck}
\Phi(\eta) =  S\sigma_d\left({\eta V_+\over \sigma_d/d}\right)^{1-1/d} + 
{1\over 2\chi}{(1 - \eta)^2V_+^2\over L^d}
\end{equation}
Here,  $0 \le \eta \le 1$, and the suggestion in \cite{BCK} is that in great generality, one can resolve a  competition
between surface and bulk energy  effects by choosing $\eta$ to minimize
$\Phi$. Defining   $C(n)$ by
\begin{equation}\label{Cdef}
C(n) = { \sigma_d  \over 2\chi S}
\left({2\over d  }\right)^2\left( {r^{d+1}_0\over  L^d  }\right) =
{2\over d\chi S} \left({\sigma_d\over d}\right)^{-1/d}\left({n+1\over 2}\right)^{(d+1)/d}L\  
\end{equation}
and  $|\Gamma_0|$  by $|\Gamma_0| = \sigma_dr_0^{d-1}$, the quantity in (\ref{fbck})
can be written as
$$\Phi(\eta) =  S|\Gamma_0|\left(\eta^{1-1/d} + C(n)(1 - \eta)^2\right)\ .$$
Notice that
$${\Phi(\eta) - \Phi(0)\over S|\Gamma_0|} = \eta( \eta^{-1/d} + C(n)\eta - 2C(n))\ .$$
By the arithmetic--geometric mean,
\begin{eqnarray}
 \eta^{-1/d} + C\eta &=& {d\over d+1}\left({d+1\over d}\eta^{-1/d}\right) +
 {1\over d+1}\left((d+1)C\eta\right)\nonumber\\
 &\ge&   \left({d+1\over d}\eta^{-1/d}\right)^{d/(d+1)}\left((d+1)C\eta\right)^{1/(d+1)}\nonumber\\
 &=&  C^{1/(d+1)}{d+1\over d^{d/(d+1)}}\ .\nonumber\\
 \end{eqnarray}
 Therefore, a minimum occurs at $\eta > 0$ if and only if 
 $$C^{1/(d+1)}{d+1\over d^{d/(d+1)}} \le 2C\ .$$
Let $C_\star$ be the value of $C$ that gives equality in this last inequality. One finds, as in \cite{BCK},
\begin{equation}\label{cstardef}
C_\star = {1\over d}\left({d+1\over 2}\right)^{(d+1)/2}\ .
\end{equation}
Moreover, with $C = C_\star$, there is equality in the application made above 
of the arithmetic geometric mean inequality if and only if $\eta^{-1/d}/d = C_\star$.  
Therefore, define $\eta_\star$ by $\eta_\star = (d C_\star)^{-d}$. One finds
\begin{equation}\label{etastardef}
\eta_\star = \left({d+1\over 2}\right)^{(d+1)/2d}\ .
\end{equation}

The heuristic argument of \cite{BCK} suggests that when $\Phi$ is minimized at $\eta = 0$, one puts all of the mass into the uniform supersaturated
state, and there is no droplet. This is the case if $C(n) < C_\star$. 
On the other hand, if $\Phi$ is minimized at a strictly positive value of $\eta$, then a strictly positive  fraction of the mass should go into a droplet.  This is the case if $C(n) > C_\star$. 
Moreover, it is easy to see that for all $C(n) > C_\star$, the minimizing value $\eta_c$ of $\eta$
satisfies $\eta_c \ge \eta_\star$. As emphasized in \cite{BCK}, this suggests that
there are never drops 
containing a volume fraction less than $\eta_\star$.  That is, at least according to this heuristic analysis, there are never droplets whose volume is less than
$$\eta_\star (\sigma_d/d)r_0^d\ .$$

The validity of this was rigorously established for the  $2$ dimensional Ising model 
in \cite{BCKb}.  We show here that the same heuristic analysis is correct for the minimization problem (\ref{mainmin}) concerning the Cahn-Hilliard free energy function ${\cal F}$. The first result concerns the value of
the ratio ${\displaystyle {f_L\over |\Gamma_0|}(n)}$ for $n =  -1 + KL^{-d/(d+1)}$, for any $K>0$,  as $L$ tends to infinity. 

\begin{thm} \label{thm1a}  For all $K>0$,
\begin{equation}\label{bckform}
\lim_{L\to \infty}  {f_L\over |\Gamma_0|}\left(-1 + KL^{-d/(d+1)}\right) = 
\inf_{0\le \eta \le 1} S\left(\eta^{1-1/d} + D(K)(1 - \eta)^2\right)
\end{equation}
where
$$D(K) = C\left(-1 + KL^{-d/(d+1)}\right)  = 
{2\over d\chi S} \left({\sigma_d\over d}\right)^{-1/d}\left({K\over 2}\right)^{(d+1)/d}$$
and $S = 2^{3/2}/3$.

Furthermore, let $K_\star$ be defined by
\begin{equation}\label{kstardef}
K_\star = 2\left({d+1\over 2}\right)^{d/2}\left({\sigma_d\over d}\right)^{1/(d+1)}
\left({\chi S\over 2}\right)^{d/(d+1)}\ .
\end{equation}
Then for all $K < K_\star$, and all $L$ sufficiently large, the infimum in (\ref{bckform}) is a minimum attained uniquely at $\eta = 0$, while for all $K > K_\star$,  and all $L$ sufficiently large, the infimum in (\ref{bckform}) is a minimum attained uniquely at $\eta = \eta_c$ where $\eta_c \ge \eta_\star$.
\end{thm}

To prove Theorem \ref{thm1a}, we prove precise upper and lower bounds on $f_L(n)$ for values of $n$ in the critical scaling regime, and from these bounds deduce (\ref{bckform}).  The remaining statements in the
theorem then follow from the discussion just above concerning the minimization of $\Phi(\eta)$.  For example, note that $K_\star$ is obtained by solving $D(K) = C_\star$ for $K$.   The upper and lower bounds on $f_L(n)$
will be presented and proved in sections 2 and 3 respectively. We conclude section 3 with the proof of Theorem
\ref{thm1a}.

The theorem suggests that the curve $n(L) = -1 + K_\star L^{-d/(d+1)}$ is critical for droplet formation,
so that for large $L$ and densities $n$ significantly below this level, the minimizers will be uniform, while
for large $L$ and densities $n$ significantly above this level, the minimizers will correspond to droplets
of a reduced radius  $\eta_c^{1/d}r_0$.   The following theorems bear this out.

\begin{thm} \label{thm1b}
For all $K < K_\star$ and $L$ sufficiently large, when
$$-1 \le n \le -1 + KL^{-d/(d+1)}\ ,$$
the unique minimizer for (\ref{mainmin}) is the uniform order parameter field $m(x) = n$. 
\end{thm}

Before stating the result concerning droplet minimizers, we must make this notion precise. To facilitate this, regard $\Omega$ as the centered cube in $\R^d$ with side length $L$ and periodic boundary conditions.

For given
$\eta$ and $n$, and hence for given $\eta$ and $r_0$, define  
a {\em sharp interface reduced radius droplet} order parameter field 
$m^\sharp_{\eta,n}(x)$
by 
$$m^\sharp_{\eta,n}(x)  =  
\begin{cases} 
\phantom{-}1  &{\rm  if}\quad  |x| < \eta^{1/d}r_0 \\ 
-1 
&{\rm  if}\quad   |x| \ge \eta^{1/d}r_0 \\
\end{cases}\ ,$$

\begin{thm} \label{thm1c}
For all  $K > K_\star$, $\epsilon>0$, and $L$ sufficiently large, when
$$-1 + KL^{-d/(d+1)} \le n \le  -1 + L^{-1/2}\ ,$$
any minimizer $m$ for  (\ref{mainmin}) is such that, after a possible translation on the torus $\Omega$,
$${1\over |r_0^d|}\int_\Omega\left|m(x) -  m^\sharp_{\eta_c,n}(x) \right|^4{\rm d}x \le \epsilon$$
where $\eta_c$ is the minimizing value of $\eta$ in (\ref{bckform}).
\end{thm}

This theorem says that for large $L$, the set on which $m$ and $m^\sharp_{\eta_c,n}$ differ by an appreciable
amount is small compared to $(\sigma_d/d)r_0^d$. In particular, on the ball where $m^\sharp_{\eta_c,n} =1$,
$m$ must be very close to $1$ on all but a negligibly small percentage of the volume of that ball. Likewise,
on the set where $m^\sharp_{\eta_c,n} =-1$,   $m$ must be very close to $-1$
on all but a set whose measure is a negligibly small percentage of the volume of the ball.   The role of the fourth power
is to make this small difference even smaller so that it is not overwhelmed by the large volume of the region external to the ball.
Any power larger than $3$ would work just as well in our argument. 

In this sense, $m$ ``looks like''  $m^\sharp_{\eta_c,n}$ for large $L$, and thus describes a droplet
of the radius predicted by the heuristic argument of \cite{BCK}.

Theorems \ref{thm1b} and \ref{thm1c} are proved in Section 4.  Finally, in Section 5, we give an explanation
for the remarkable efficacy of the simple trial function used in Section 2. In particular, we see which features of
the free energy functional (\ref{(1)}) are responsible for this. The point is that for the free energy functionals coming from models with a non local interaction, such as the ones considered in \cite{CCELM},  as well as in \cite{ABCP} and \cite{BBP}, these features are not present. However, the analysis in Section 5 leads to a method for constructing trial functions of high accuracy
that does apply to such cases, as well as to  (\ref{(1)}).

\medskip
\centerline{\bf Acknowledgements}
\medskip

This work was completed during January through April of 2005 when the authors were visitors at
 I.H.\'E.S.   The authors thank
Professor J.P. Bourguignon for kind hospitality at the I.H.\'E.S.  This work was first presented at the P.D.E. seminar at E.N.S., Paris. We thank the participants of the seminar, particularly Benoit Perthame, for questions that
have  helped improve the written presentation.
\medskip

\section{The upper bound}\label{upper}

\subsection{The interpolating family of trial functions} \label{ub1}

For $0 \le \eta \le 1$, let $r_\eta = \eta^{1/d}r_0$ be the radius of a ball whose volume is $\eta$ times the volume of
a ball with the equimolar radius, $r_0$.  The arguments of Biskup, Chayes and Kotecky suggest that one should use as a
trial function a function of the form
\begin{equation}\label{partial}
m_{\eta{\rm dr}}(x) =  m_0(|x| - r_\eta) +  \alpha(\eta)\ ,
\end{equation}
where  $m_0$ is a transition profile that very nearly minimizes the cost in free energy of making
the transition from $m = +1$ to $m=-1$,  and $\alpha(\eta)$ is a constant determined by the constraint $\int_\Omega m_{\eta{\rm dr}}(x){\rm d}x = n|\Omega|$.  As in Section 1, we are taking $\Omega$ to be the centered cube in $\R^d$ with side length $L$.  

As $\eta$ varies in the interval  $0 < \eta < 1$, the family of ``fractional droplet'' trial functions
defined in (\ref{partial}) interpolates between $m_{\rm uni}$, for $\eta=0$ and    $m_{\rm emd}$, for $\eta =1$.
Of course, it remains to choose $m_0$. 

\subsection{Planar surface tension and the choice of $m_0$} \label{ub2}

The natural choice for $m_0$ is given by considering the  problem of minimizing the cost per unit area 
in free energy of an infinite planar interface between the $+1$ and $_1$ phase. Denote this quantity by $S$; it will turn out to be the same constant $S$ that appears in (\ref{surften}). That is,
$$S  = \inf\left\{\ \int_{\Bbb R}  \left({1\over 2}|m'(z)|^2 +  F(m(z))\right){\rm d}z \ :\ 
\lim_{z\to \pm \infty}m(z) = \mp 1\ \right\}\ .$$

Let $\bar m$ denote  minimizer for this variational problem with $\bar m(0) =0$.
The Euler--Lagrange equation satisfied by $\bar m$ is 
$\bar m''(z) = F'(\bar m(z))$.
Multiplying both sides by $\bar m'(x)$, and integrating from $-\infty$ to $z$, we obtain
\begin{equation}\label{equal}
(\bar m'(z))^2 = 2F(\bar m(z))\ ,
\end{equation}
since $\lim_{z\to -\infty} m'(z) =  \lim_{z\to -\infty} F(m(z)) = 0$.

One now easily deduces that $\bar m(z) = -{\rm tanh}(z/\sqrt{2})$, from which one could compute $S$. However, there is anther route that is more informative and useful in what follows:
From (\ref{equal}), we see that
$$S = 2\int_{-\infty}^\infty F(\bar m(z)){\rm d}z\ ,$$
and furthermore,
\begin{eqnarray}
\int_{-\infty}^\infty F(m(x)){\rm d}x &=&  \int_{-\infty}^\infty {F(m(x))\over m'(x)} m'(x){\rm d}x\nonumber\\
&=&  \int_{-\infty}^\infty -\sqrt{F(m(x))\over 2} m'(x){\rm d}x\nonumber\\
&= & \int_{-1}^1 \sqrt{F(h)\over 2} {\rm d}h\nonumber\\
\end{eqnarray}
Thus,
\begin{equation}\label{intform}
S = \int_{-1}^1 \sqrt{2 F(h)} {\rm d}h\ ,
\end{equation}
and hence  $S = 2^{3/2}/3$, which provides the numerical value
for $S$ that is quoted in Theorem \ref{thm1a}. However, in what follows, it is the integral formula, and not so much the numerical value, that turns out to matter.

We  now choose $m_0$. We cannot simply choose $m_0 = \bar m$ since then
$m_0(|x| - r_\eta)$ would not define a smooth, or even continuous,  function on $\Omega$. 
However, only mild modification are required. Since we are interested in values of $r$
with $r = {\cal O}\left(L^{d/(d+1)}\right)$, define
$$m_0(z) =  
\begin{cases} 
\phantom{-} \bar m(z)  &{\rm  if}\quad  |z| < L^{(d-1)/(d+1)} \\ -{\rm sgn}(z)  
&{\rm  if}\quad   |z| > 2L^{(d-1)/(d+1)}  \\
\end{cases}\ ,$$
and smoothly interpolate in such a way that $m_0$, like $\bar m$, is odd. 
With any such interpolation, $m_0(|x| - r_\eta)$ defined a smooth function on $\Omega$, and the difference between $m_0$ and $\bar m$ goes to zero exponentially fast as $L$ tends to infinity.

\subsection{The determination of $\alpha(\eta)$} \label{ub3}

The constraint equation is
$$\int_\Omega m_{\eta{\rm dr}}(x){\rm d}x = nL^d =   \left[2(\sigma_d/d)r^d_0 -L^d\right] $$
and hence
\begin{equation}\label{alphaform}
\alpha(\eta) L^d =   \left[2(\sigma_d/d)r^d_0 -L^d\right]  - 
\int_\Omega m_0(|x| - r_\eta){\rm d}x\ .
\end{equation}
We require sharp estimates on the integral on the right.

\medskip

\begin{lm} \label{intbnd} Define the constant $M$ by
$$M =  \int_\R ({\rm sgn}(z) - {\rm tanh}(z/\sqrt{2}))z{\rm d}z\ .$$
For all $L^{(d-1)/(d+1)} < r_\eta < r_0$,  
$$\int_\Omega m_0(|x| - r_\eta){\rm d}x  = \left[2(\sigma_d/d)r_0^d \eta  -L^d\right]  - (d-1)M\sigma_d r_0^{d-2}\eta^{(d-2)/d} +
{\cal O}\left(e^{-L^{1/4}}\right)$$
when $d=2$ or $d=3$. For higher dimension, the only difference is that the error term is
${\cal O}(r_0^{d-4})$.
\end{lm} 

\medskip

\noindent{\bf Proof:}  Note that
$$\int_\Omega m_0(|x| - r_\eta){\rm d}x =  \left[2(\sigma_d/d)r_\eta^d -L^d\right]    - 
\int_\Omega ({\rm sgn}(|x| - r_\eta) + m_0(|x| - r_\eta)) {\rm d}x\ .$$
Define
$$I_1 =  \int_{|x| \le 2r_\eta}({\rm sgn}(|x| - r_\eta) + m_0(|x| - r_\eta)) {\rm d}x \qquad{\rm and}\qquad
I_2 =  \int_{|x| > 2r_\eta}({\rm sgn}(|x| - r_\eta) + m_0(|x| - r_\eta)) {\rm d}x\ .$$
We easily see that for all dimensions $d$,
$I_2 = {\cal O}(e^{-L^{1/4}})$.
Moreover, using polar coordinates,
$$I_1 = \sigma_d \int_0^{2r_\eta}({\rm sgn}(s - r_\eta) + m_0(s - r_\eta)) s^{d-1}{\rm d}s\ .$$
Introducing the new variable $z =s-r_\eta$, we see that if we extend the integration in $z$ over the whole real line, we only make an error of size ${\cal O}(e^{-L^{1/4}})$ at most, and so
$$I_1 =  \sigma_d r^{d-1}\int_\R ({\rm sgn}(z) + m_0(z)) \left(1 - {z\over r_\eta}\right)^{d-1}{\rm d}z +
{\cal O}(e^{-L^{1/4}})\ .$$
Taking into account the fact that $({\rm sgn}(z) + m_0(z))$ is odd and rapidly decaying, we see that for $d=2$ or $d=3$,
$$\int_\R ({\rm sgn}(z) + m_0(z)) \left(1 - {z\over r_\eta}\right)^{d-1}{\rm d}z = {d-1\over r_\eta}
\int_\R ({\rm sgn}(z) + m_0(z))z{\rm d}z\ .$$
In any dimension, this gives the leading order correction. 
This, together with the definition of $m_0(z)$
in terms of $\bar m(z) = -{\rm tanh}(z/\sqrt{2})$,  yields the result. \qed
\medskip

Therefore, (\ref{alphaform}) 
together with Lemma \ref{intbnd} yield for $d=2$ or $d=3$ that
\begin{equation}\label{alphaexp}
\alpha(\eta) =   2(\sigma_d/d){r^d_0\over L^d}(1-\eta) +  (d-1)M\sigma_d {r_0^{d-2}\over L^d}\eta^{(d-2)/d}+ 
{\cal O}(e^{-L^{1/4}})\ .
\end{equation}
The only difference for higher dimensions $d$ is that ${\cal O}(e^{-L^{1/4}})$ must be replaced by
${\cal O}(r_0^{d-4}/L^d)$.  Notice that unless $\eta =1$, the first explicit correction is already very small compared to the leading term;
it is smaller by a factor of   $r_0^{-2}$.   In the critical regime, by (\ref{critreg}), $r_0^{-2} \asymp L^{-2d/d+1)}$.
Moreover, we see that in the critical scaling regime, except when $\eta =1$,
\begin{equation}\label{a2}
\alpha(\eta) \asymp L^{-d/(d+1)}\ .
\end{equation}

\subsection{Computation of ${\cal F}(m_{\eta{\rm dr}})$} \label{ub4}

With the trial function specified, we now compute ${\cal F}(m_{\eta{\rm dr}})$.

\medskip

\begin{lm} \label{fbnd}  In the critical scaling regime $r_0 \asymp L^{d/(d+1)}$,    
\begin{equation}\label{upbnd}
{\cal F}(m_{\eta{\rm dr}})  \le  \Phi(\eta)   -8(\sigma_d/d)^3{r_0^{3d}\over L^{2d}}(1-\eta)^3 +
{\cal O}\left(L^{(d^2-3d)/(d+1)}\right)\ ,
\end{equation}
where the first term on the right is ${\cal O}\left(L^{(d^2-d)/(d+1)}\right)$ and the second is
${\cal O}\left(L^{(d^2-2d)/(d+1)}\right)$.
\end{lm} 

\medskip

Notice that the leading term in the upper bound is exactly $\Phi(\eta)$, and that the next term is negative.

\medskip

\noindent{\bf Proof:}   To simplify the notation, we write $m_0$ to denote $m_0(|x| - r_\eta)$ and $\alpha$
to denote $\alpha(\eta)$ 
so that $m_{\eta{\rm dr}} = m_0 + \alpha$. Then
$$F(m_{\eta{\rm dr}}) = F(m_0) + F'(m_0)\alpha + {1\over 2}F''(m_0)\alpha^2 + 
{1\over 6}F'''(m_0)\alpha^3  + {1\over 4}\alpha^4\ .$$

We are required to produce a close upper bound on the integral of each of these terms over $\Omega$.
We start with $\alpha \int_\Omega F'(m_0){\rm d}x$.

Note that $F'(m) = m^3 -m$, Since $m_0^3(z) - m_0(z)$ is an odd, rapidly decaying function of $z$, estimates just like the ones employed in the proof of Lemma (\ref{intbnd})
show that 
$$\int_\Omega  F'(m_0) {\rm d} x = \sigma_dr_\eta^{d-1}\int_\R (\bar m^3(z) - \bar m(z))
\left(1 - {z\over r_\eta}\right)^{d-1}{\rm d}z + {\cal O}(e^{-L^{1/4}})\ .$$
Then, with the constant $B$ defined by
$$B =  \int_\R (\bar m^3(z) - \bar m(z))z{\rm d}z\ ,$$
we have for $d=2$ or $d=3$ that
\begin{equation}\label{fp1}
\int_\Omega  F'(m_0) {\rm d} x = \sigma_d r_0^{d-2}B \eta^{(d-2)/d} + {\cal O}(e^{-L^{1/4}})\ ,
\end{equation}
and the same is true for $d \ge 4$ except that the error term must be replaced by ${\cal O}(r_0^{d-4})$.

Next, $F''(m_0) = 3m_0^2 -1 \le 2 = 1/\chi$.  Therefore 
\begin{equation}\label{fp2}
\int_\Omega{1\over 2}F''(m_0){\rm d}x  \le {1\over 2\chi} L^d\ .
\end{equation}

Finally, $F'''(m) = 6m$, and so
\begin{equation}\label{fp3}
\int_\Omega {1\over 6}F'''(m_0){\rm d}x = \int_\Omega m_0{\rm d}x\ ,
\end{equation}
and this integral has been computed in Lemma \ref{intbnd}.

Now let $I$ denote the integral
$$I = \int_\Omega\left[F'(m_0)\alpha + {1\over 6}F'''(m_0)\alpha^3  + {1\over 4}\alpha^4\right]{\rm d}x\ .$$
In the critical scaling regime, the dominant contribution to $I$ comes from the $F'''$ term, and is
$-L^d\alpha^3 \asymp L^{(d^2-2d)/(d+1)}$. Each of the terms in the integrand contributes at the order
$L^{(d^2-3d)/(d+1)}$ in the critical scaling regime, and we have
$$I = -L^d\alpha^3  +  \left[\sigma_d r_0^{d-2}B \eta^{(d-2)/d} \alpha + 2(\sigma_d/d)r_0^d\eta\alpha^3 
+ \alpha^4L^d/4\right] + {\cal O}\left(L^{(d^2-4d)/(d+1)}\right)\ .$$
Using (\ref{alphaform}), we can express this as
\begin{eqnarray}\label{Ibnd}
I &=& -8(\sigma_d/d)^3{r_0^{3d}\over L^{2d}}(1-\eta)^3 + \left[
2d(\sigma_d/d)^2B{r_0^{2d-2}\over L^d}\eta^{2-2/d}
+  4(\sigma_d/d)^4{r_0^{4d}\over L^{3d}}(1-\eta)^3(1+3\eta)\right]\nonumber\\
&+& {\cal O}\left(L^{(d^2-4d)/(d+1)}\right)\ ,\nonumber\\
\end{eqnarray}
where the first term on the right is proportional to $L^{(d^2-2d)/(d+1)}$, and the second 
is proportional to $L^{(d^2-3d)/(d+1)}$.

Finally, we have to estimate $\int_\Omega \left[ |\nabla m_0|^2 + F(m_0)\right]{\rm d}x$. 
Once more, estimates just
like the ones employed in the proof of Lemma \ref{intbnd}
show that 
$$\int_\Omega \left[ |\nabla m_0|^2 + F(m_0)\right]{\rm d}x  \approx \sigma_dr_\eta^{d-1}
\int_\R\left[ |\bar m'(z)|^2 + F(\bar m(z))\right](1 + z/r_\eta)^{d-1}{\rm d}z$$
where the errors are exponentially small in $L^{1/4}$. But because $m_0$ is so close to $\bar m$,
this only differs from  $S|\Gamma_0|\eta^{1-1/d}$ by errors that are ${\cal O}(r_0^{d-3})$.  
In the asymptotic scaling regime, $r_0^{d-3} \asymp L^{(d^2-3d)/(d+1)}$.

Combining estimates, we have
\begin{eqnarray}
{\cal F}(m_{\eta{\rm dr}}) 
&\le &  S\sigma_d r^{d-1}  + 
{1\over 2\chi}\left({2\sigma_d\over d}\left({r_0\over L}\right)^d - {2\sigma_d\over d}\left({r\over L}\right)^d\right)^2L^d + {\cal O}(1)\nonumber\\
&=& S\sigma_d r_0^{d-1}\left({r\over r_0}\right)^{d-1}  + 
{ L^d  \over 2\chi }
\left({2\sigma_d r^d_0\over d L^d    }\right)^2\left(1 - \left({r\over r_0}\right)^d
\right)^2+ {\cal O}(1) \nonumber\\
&=&  S|\Gamma_0|(\eta^{1/d} + C(n)(1-\eta)^2) +  {\cal O}\left(L^{(d^2-2d)/(d+1)}\right)\ . \nonumber\\
\end{eqnarray}
\qed

\section{The lower bound}\label{lower}

\subsection{An {\em A priori} pointwise upper bound} 

Standard compactness arguments show that the infimum in (\ref{mainmin}) is attained at a minimizer $m(x)$
which satisfies the Euler--Lagrange equation
\begin{equation}\label{el}
-\Delta m(x) + m^3(x) - m(x) + \mu = 0\ ,
\end{equation}
where $\mu$ is a Lagrange multiplier corresponding to the constraint in (\ref{mainmin}).

Our immediate goal is to prove an {\em a priori} pointwise upper bound on a minimizer $m$ that is very close to $1$
in the critical scaling regime. Such a bound can be obtained from
 the Euler--Lagrange equation and the maximum principle.
 
Let $x_{\rm min}$
and $x_{\rm max}$ be such that for all $x$,
$$m( x_{\rm min}) \le m(x) \le m(x_{\rm max})\ .$$
These exist since any solution of the Euler--Lagrange equation is continuous.

We will now show that  $m(x_{\rm max})$ cannot be too large. Define numbers $\lambda$ and $\nu$ by
\begin{equation}\label{ln}
1+ \lambda =   m(x_{\rm max})  \qquad {\rm and}\qquad   -1+ \nu =   m(x_{\rm min}) \ .
\end{equation}
It will be convenient in the arguments leading to the proof to  write $n$ in the form 
\begin{equation}\label{deldef}
n = -1 +\delta\ .
\end{equation}
Notice that in the critical scaling regime, $\delta \asymp L^{-d/(d+1)}$. Also, from (\ref{Aofn}) and (\ref{equimolar}),
\begin{equation}\label{deltaequi}
\delta = 2{V_+\over L^d} = 2(\sigma_d/d){r_0^d\over L^d}\ .
\end{equation}

\medskip
\begin{lm} \label{lb1}    For any solution of the Euler--Lagrange equation (\ref{el}), let $\lambda$ and $\nu$
be given by (\ref{ln}). Then
$\nu \ge \lambda$. 
Consequently, if $m$ is any minimizer for (1), 
\begin{equation}\label{top}
\delta \ge \lambda\ .
\end{equation}
\end{lm}
\medskip

\noindent{\bf Proof: } Evidently, $\Delta m(x_{\rm max}) \le 0$, and so from (\ref{el})  and (\ref{ln}),
$(1+\lambda)^3- (1+\lambda) + \mu \le 0$, or
$$2\lambda + 3\lambda^2 + \lambda^3   \le -\mu$$
In the same way, from (\ref{el})  and (\ref{ln}) we have
$$2\nu - 3\nu^2 + \nu^3 + \mu \ge  -\mu\ .$$
Therefore,
$$2\nu - 3\nu^2 + \nu^3 \ge   2\lambda + 3\lambda^2 + \lambda^3\ .$$
It evidently follows that
$$2\nu  + \nu^3 \ge   2\lambda  + \lambda^3\ ,$$
and since $f(x) = 2x + x^3$ is monotone increasing, it follows that $\nu \ge \lambda$. 

Next, since the average value of any function is no less than its minimum, it follows that 
$n \ge -1+ \nu$, and by (4), this means $\delta \ge \nu$. Combining estimates, we have (\ref{top}).
\qed

\medskip

\subsection{An {\em A priori} lower bound on $m(x_{\rm max})$}

We next show that any for  non constant minimizer $m$, it cannot be that $m(x_{\rm max})$ is much smaller than
$1$.  For this purpose,
define $w$ by $w(x) = m(x)-n$. For $m$ satisfying the constraint in (\ref{mainmin}),
\begin{equation}\label{wcon}
\int_\Omega w(x){\rm d}x = 0\ .
\end{equation}

Clearly, $\int_\Omega |\nabla m|^2{\rm d}x = \int_\Omega |\nabla w|^2{\rm d}x$, and
$$
\int_\Omega{1\over 4}(m^2 -1)^2{\rm d}x = L^2(n^2 -1)^2 + 
\int_\Omega{1\over 4}\left({1\over 2}(3n^2 -1)w^2 + nw^3 + {1\over 4}w^4\right){\rm d}x\ ,$$
since the terms linear in $w$ drop out due to (5b), and
$$nw^3 + {1\over 4}w^4 = {1\over 4}w^2(w+2n)^2 - n^2w^2\ .$$
Hence, if we define the functional ${\cal G}$ by
\begin{equation}\label{calGdef}
{\cal G}(w) =  {1\over 2}\int_\Omega|\nabla w|^2{\rm d}x + \int_\Omega G(w){\rm d}x\ ,
\end{equation}
where
\begin{equation}\label{Gdef}
G(w) = 
{n^2 -1\over 2 }w^2 
+ {1\over 4} w^2(w+2n)^2\ ,
\end{equation}
we have
\begin{equation}\label{FtoG}
{\cal F}(m) = {\cal F}(n) + {\cal G}(w)
\end{equation}
whenever $m$ satisfies the constraint in (\ref{mainmin}).

\medskip
\begin{lm} \label{lb2}  Let $m$ be any minimizer for (\ref{mainmin}), and suppose that $m$ is not constant.
Then 
\begin{equation}\label{top2}
m(x_{\rm max}) \ge 1 -  \delta - 2\sqrt{\delta}\sqrt{1 - \delta/2}\ .
\end{equation}
\end{lm}
\medskip

\noindent{\bf Proof:} Notice that 
$${n^2 -1\over 2}
+ {1\over 4}(w+2n)^2 < 0$$
if and only if 
$$z_-  < w <  z_+\ $$
where
$$z_\pm = -2n \pm \sqrt{2-2n^2}  = 2 -2\delta \pm 2\sqrt{\delta}\sqrt{1 - \delta/2}\ .$$
Since $m$ is not constant $\int_\Omega |\nabla w|^2{\rm d}x > 0$. 
Thus ${\cal G}(w) > 0$ unless $w(x_{\rm max}) \ge 
2 -2\delta - 2\sqrt{\delta}\sqrt{1 - \delta/2}$. If $m$ is a minimizer,  ${\cal G}(w) > 0$ is impossible,
on account of (\ref{FtoG}).  Since $m(x_{\rm max}) = n + w(x_{\rm max})$, we have the estimate.
\qed

\medskip

\subsection{A partition of $\Omega$} 

We now partition $\Omega$ in to three pieces, one of which will contribute a surface tension term to the free energy, another of which will contribute a compressibility term, and another that will be negligible.

Returning to the original dependent variable $m$, we fix a number $\kappa>0$ to be determined below.
However, to fix our ideas for the time being, suppose that $\kappa = {\cal O}(\delta^{1/3})$. 
Define numbers $h_+$ and $h_-$ by
\begin{equation}\label{hpmdef}
h_+ = 1-\kappa\qquad{\rm and}\qquad h_-= -1 + \kappa\ .
\end{equation}
Define the sets $A$, $B$ and $C$ by
$$A = \{\ x\ :\ h_- \le m(x) \le h_+\ \}\qquad B =  \{\ x\ :\ m(x) \le h_-\ \}\ ,$$
and
$$C =  \{\ x\ :\  m(x) \ge h_+\ \}\ .$$
If $m$ is a non constant minimizer, and  $\kappa = {\cal O}(\delta^{1/3})$, then for $L$ large enough, $C$
will be non empty by Lemma 3.2. Define a radius $R$ by
\begin{equation}\label{Rdef}
(\sigma_d/d) R^d =  |C|\ ,
\end{equation}
where the right hand side denotes the measure of $C$.   Evidently $R$ is the radius of the ball with 
the same volume as $C$. 

We shall obtain a lower bound on $f_L(n)$ by separately estimating the integrals
\begin{equation}\label{ABC}
I_A = 
\int_A\left[{1\over 2} |\nabla m|^2 + F(m)\right]{\rm d}x \qquad {\rm and}\qquad
I_B = 
\int_B\left[{1\over 2} |\nabla m|^2 + F(m)\right]{\rm d}x\ .
\end{equation}

\subsection{The surface tension contribution} 

We now prove a lower bound on $I_A$, which corresponds to the surface tension contribution to the free energy. 
The lower bound  is obtained through use of the co--area formula \cite{AL}, \cite{Fed}, which
expresses the volume element in $\Omega$ as
$${\rm d}x = {1\over |\nabla m(x)|} {\rm d}\sigma_h{\rm d}h\ ,$$
where ${\rm d}\sigma_h$ is the surface area along $\Gamma_h$, the level set $\{m(x) = h\}$, or, more properly put, the $d-1$ dimensional Huasdorf measure on this set.

It is worth noting at this point that the rearrangement inequalities of the sort discussed in \cite{CCELM}
apply in this case, and allow us to conclude that for any minimizer $m$, the level sets are {\em symmetric monotone}.  If we translate so that the maximum of $m$ is at $0$, then this means that for any $h$, and any of the standard basis vectors $\vec e_j$, $j=1,\dots, d$, the set of $t$ for which
$m(t\vec e_j) > h$ is a symmetric interval.  In particular, $\Gamma_h$ is a rectifiable, simply connected curve.

In applying the co--area formula, we shall gloss over certain standard technical issues. These are all explained, for example,  in the discussion of the Faber--Krahn inequality in \cite{Bu} or \cite{CK}, where the co--area formula is applied to another variational problem, namely the one for the fundamental eigenvalue for the Laplacian in a a domain in $\R^d$. Those readers who are not familiar with the use of the co--area formula in proving inequalities
such as the Faber--Krahn inequality may wish to consult the references cited above. For those who are, we proceed with the proof.

\medskip
\begin{lm}\label{lb3} Let $m$ be any non constant minimizer for (\ref{mainmin}), 
and suppose that $\kappa = {\cal O}(\delta^{1/3})$. Then for $L$ large enough,
Then 
\begin{equation}\label{IAbnd}
 I_A  \ge  (\sigma_d)R^{d-1}\left(S -2\kappa \right)\ .
 \end{equation}
\end{lm}
\medskip
\noindent{\bf Proof:} 
By the co--area formula, 
$$I_A = \int_{h_-}^{h_+}\int_{\Gamma_h} 
\left({1\over 2}|\nabla m(x)| + {F(h)\over |\nabla m(x)|}\right){\rm d}\sigma_h{\rm d}h\ .$$
By the arithmetic--geometric mean inequality,
$${1\over 2}\left(|\nabla m(x)| + 2{F(h)\over |\nabla m(x)|}\right) \ge  \sqrt{2F(h)}\ ,$$
and therefore,
$$I_A \ge 
\int_{h_-}^{h_+}|{\Gamma_h}| \sqrt{2F(h)}{\rm d}h$$
where $|{\Gamma_h}|$ denote the one dimensional Haussdorf measure of $\Gamma_h$. 

Note that $\Gamma_h$ encloses a region whose volume is at least $|C|$. By the isoperimetric inequality on
the torus,  the length of the boundary of such a region is at least
$\sigma_d^{1/d}(dV)^{1-1/d}$ provided $|C| \le (2^{d/(d+1)}L^2)/(d\sigma_d^{d-1})$, and is at least $2L^{d-1}$ otherwise. 
( If $U$ is any domain in $\Omega$ with $|U| \le (2^{d/(d+1)}L^2)/(d\sigma_d^{d-1})$, then the surface area of the boundary of $U$ is no less than that of
 a ball of volume $|U|$, namely $\sigma_d^{1/d}(dV)^{1-1/d}$). Otherwise, the lower bound on the perimeter is simply
$2L^{d-1}$, and this is achieved by a ``strip'' of width $|U|/L^{d-1}$ in the torus.) Therefore, defining $P$ by
$$P = \min\{\  \sigma_d^{1/d}(d|C|)^{1-1/d} \ ,\ 2L^{d-1}\ \}\ ,$$
we have
$|\Gamma_h| \ge P$ for all $h_- \le h \le h_+$. Hence

$$I_A  \ge 
P \int_{h_-}^{h_+}  \sqrt{2F(h)}{\rm d}h$$
By (\ref{intform}), this yields
$$I_A  \ge 
P\left(S - \int_{-1}^{h_-}  \sqrt{2F(h)}{\rm d}h - \int_{h_+}^1 \sqrt{2F(h)}{\rm d}h\right)$$

Furthermore,
$$\sqrt{2F(h)} =(1 -h^2)/  \sqrt{2}  \le  1\ , $$
and hence
$\int_{-1}^{h_-}  \sqrt{2F(h)}{\rm d}h \le \kappa$ and 
$\int_{h_+}^1 \sqrt{2F(h)}{\rm d}h \le \kappa$.  This gives us $I_A \ge P(S -2\kappa)$. Now if $|C| \ge L^2/\pi$,
this would imply $I_A \ge 2L(S - \kappa)$, which is much larger than ${\cal F}(n)$. This is therefore impossible when $m$ is a minimizer for (\ref{mainmin}), and so with $R$ defined by (\ref{Rdef}), we have the bound (\ref{IAbnd}).

\medskip

\subsection{The bulk contribution} 

In this subsection, we prove a lower bound on the contribution to the free energy from $B$. For this purpose, we first require an upper bound on $|A|$ which shows that, for large $L$, $|A|$ is negligible compared to $|C|$. 
Ideally, one might hope that $A$ is an annular region about $C$, and to obtain a ``surface term'' type bound for $A$, showing that is
is bounded by a multiple of $r_0^{d-1}\asymp L^{d/(d+1)}$. However, as one can see from the proof of Lemma \ref{lb3}, even if $C$ were spherical and $A$ was an annulus about it, one would have to take
the annulus to be fairly ``thick'' in order to capture most of $S$ in the estimate (\ref{IAbnd}). Thus, the following simple estimate is rather sharp.

\medskip
\begin{lm}\label{lb4}
Let $m$ be any minimizer for (\ref{mainmin}). Then 
$$|A| \le 2F(n){L^d\over \kappa^2} \le  2 {\delta^2\over \kappa^2}L^d\ .$$
\end{lm}
\medskip

\noindent{\bf Proof:}   Since
$$F(h_+) = \kappa^2(1- \kappa/2)^2  =
F(h_-)\ ,$$
it is easy to see that  uniformly on $A$, 
$$F(m(x)) \ge \kappa^2(1- \kappa/2)^2\ \eqno(12)$$
Therefore
$$I_A \ge |A|\kappa^2(1- \kappa/2)^2\ .$$
On the other hand, since $m$ is a minimizer, 
$$I_A < {\cal F}(n) = F(n)L^2\ .$$
In the range of $\delta$ being considered, $(1- \kappa/2)^2 \ge 1/2$. \qed

\medskip

Our next goal is a lower bound on $I_B$. 
Notice that on $(-\infty,h_-)$, $F$ is strictly convex. In fact, $F''(h) \ge 3h_-^2 -1$.
Define the quantity $\chi_-$ by
$${1\over \chi_-} = F''(h_-) = 3h_-^2 -1\ .$$
Then, by Taylor's Theorem, and using the fact that $F(-1) = F'(-1) = 0$, we have
$$F(m(x)) \ge {1\over 2 \chi_-}(m(x) +1)^2$$
everywhere on $\{m \le h_-\}$. 

Therefore,
\begin{eqnarray}
\int_B F(m(x)){\rm d}x &=&   |B|\left({1\over |B|}
\int_{B} F(m(x)){\rm d}x\right)\nonumber\\
&\ge&
 |B|{1\over 2\chi_-}\left({1\over |B|}
\int_{B} (m(x)+1)^2{\rm d}x\right)\nonumber\\
&\ge& 
 |B|{1\over 2\chi_-}\left({1\over |B|}
\int_{B} (m(x)+1){\rm d}x\right)^2\nonumber\\
&=& {1\over 2\chi_- |B|}\left(
\int_{B} m(x){\rm d}x + |B|\right)^2\ .\nonumber\\
\end{eqnarray}

We now need an upper bound and lower bounds on $|B|$ and 
$\int_{B} m(x){\rm d}x$.
Note that
$$
\int_{B} m(x){\rm d}x = nL^d  - \int_{C} m(x){\rm d}x - \int_{A} m(x){\rm d}x\ .$$
By Lemma \ref{lb1} and the definition of $R$,
$$(1-\kappa)(\sigma_d/d) R^d \le    \int_{C} m(x){\rm d}x   \le   (1+\delta)(\sigma_d/d) R^d\ .$$
By Lemma \ref{lb4}, 
$$-|A| \le h_-|A| \le   \int_{A} m(x){\rm d}x  \le h_+|A| \le |A|\ .$$
Thus since $\kappa > \delta$,
$$\left| \int_{B} m(x){\rm d}x - (nL^d - (\sigma_d/d) R^d)\right| \le |A| + \kappa (\sigma_d/d) R^d\ .$$
Next, it is evident that $|B| = L^d - (\sigma_d/d) R^d - |A|$.  Therefore
$$\left| \left(\int_{B} m(x){\rm d}x + |B|\right) - (\delta L^2 - 2(\sigma_d/d) R^d)\right| \le 2|A| + 
\kappa (\sigma_d/d) R^d\ .$$
Hence, if we define $\epsilon$ by
$$\epsilon = 2(2|A|/L^d + \kappa (\sigma_d/d) (R/L)^d)|\delta  - 2(\sigma_d/d) (R/L)^d|\ ,$$
we have
$$\left(\int_{B} m(x){\rm d}x + |B|\right)^2 \ge (\delta L^d - 2(\sigma_d/d) R^d)^2 - \epsilon L^{2d} \ .$$
By Lemma \ref{lb4}, $|A|/L^2 = {\cal O}(\delta^2/\kappa^2)$, and for any minimizer, we must have $R = {\cal O}(L^{d/(d+1)}$, since otherwise, if $R$ were any larger, the contribution from the interface as estimated in Lemma \ref{lb3}
would already exceed the free energy for the uniform trial function.  Thus, $(R/L)^d = {\cal O}(L^{-d/(d+1)}) = 
{\cal O}(\delta)$.  Therefore,
$$\epsilon ={ \cal O}\left({\delta^3\over \kappa^2} + \kappa\delta^2\right)\ .$$
With the choice $\kappa = \delta^{1/3}$, this gives us
\begin{equation}\label{order}
\epsilon = {\cal O}(\delta^{7/3})\ ,
\end{equation}
with the essential point being that this is negligible compared to $\delta^2$ as $L$ tends to infinity in the critical scaling regime.

Finally, since $|B| < L^d$, this proves the following bound:
\begin{lm}\label{lb5}
 Let $m$ be any non constant minimizer for (\ref{mainmin}). Then, with $\epsilon$ given as above,
$$I_B \ge  {L^d\over 2\chi_- }\left((\delta  - 2(\sigma_d/d)  (R/L)^d)^2 - \epsilon\right) \ .$$
\end{lm}

\medskip

It now follows from Lemmas \ref{lb3} and \ref{lb5}   that for any non constant minimizer $m$,
\begin{eqnarray}
{\cal F}(m) &\ge& I_A + I_B\nonumber\\
&\ge&   \sigma_d R^{d-1}\left(S -2\kappa \right) +
{L^d\over 2\chi_-}(\delta  - 2(\sigma_d/d)(R/L)^d)^2 - {L^d\over 2\chi_-}\epsilon\ .\nonumber\\
\end{eqnarray}
It now remains to optimize this over $R$. 

Now introduce $S_- = \left(S -2\kappa \right)$ and $\eta = R^d/r_0^d$. Then we can rewrite this lower bound as
\begin{equation}\label{goodform}
{\cal F}(m)  \ge S_-|\Gamma_0|\left(\eta^{1-1/d} + {S\chi\over S_-\chi_-}C(n)(1-\eta)^2 \right) - 
{L^d\over 2\chi_-}\epsilon\ .
\end{equation}

\medskip

\noindent{\bf Proof of Theorem \ref{thm1a}}:   
As $L\to \infty$ in the critical scaling regime, $S_- \to S$ and  $\chi_- \to \chi$. Moreover, from (\ref{order}),  (\ref{deltaequi}) and the definition of $\epsilon$,
$L^d\epsilon/|\Gamma_0| \to 0$ as $L\to \infty$. Thus, (\ref{goodform}) provides the lower bound
needed to prove (\ref{bckform}). The upper bound is provided by Lemma \ref{ub2}.  The remaining statements
follow from the analysis of the minimization of the phenomenological free energy function (\ref{fbck}) that was explained in the introduction. \qed

\section{The structure of the minimizers}\label{main}

\subsection{The proof of Theorem \ref{thm1b}} 

Suppose that $n = -1 + KL^{d/(d+1)}$ where $K < K_\star$. We would like to conclude from (\ref{goodform})
that any non constant trial function $m$ has a higher free energy than the uniform trial function $m(x) = n$, at least for all sufficiently large $L$.

Recalling that $S|\Gamma_0|C(n) = {\cal F}(n)$, define $\bar\eta$ by
$$\bar\eta = \sup\left\{ \eta \ :\ 
S_-|\Gamma_0|\left(\eta^{1-1/d} + {S\chi\over S_-\chi_-}C(n)(1-\eta)^2 \right) - 
{L^d\over 2\chi_-}\epsilon <  S|\Gamma_0|C(n) \ \right\}$$

As in the proof of Theorem \ref{thm1a}, for all $L$ sufficiently large, $S_-$ is sufficiently close to $S$, and $\chi_-$ is sufficiently close to $\chi$ that 
$${S\chi \over S_-\chi_-}C(n) < C$$
for some $C < C_\star$. For $C < C_\star$, the unique minimizer of 
$$\eta \mapsto \eta^{1-1/d} + C(1-\eta)^2$$
is $\eta =0$. Therefore, since $\epsilon L^d/|\Gamma_0|\to 0$ as $L\to \infty$, it follows that
$\bar\eta \to 0$ as $L\to \infty$. 

Now, as in the previous section, for any non uniform minimizer $m$, there is a relation between $\eta$
and the size of the level set  $|\{m > 1- \kappa\}|$ given by
$\eta = (R/r_0)^{1/d}$ and 
$|\{m > 1- \kappa\}| = (\sigma_d/d)R^d$.  Here, as in the last section, $\kappa = \delta^{1/3}$ 
with $\delta$ given by (\ref{deldef}).  It follows from (\ref{goodform}) and the definition of $\bar\eta$
that for any non constant minimizer $m$, $\eta < \bar\eta$, and so
$|\{m > 1- \kappa\}|$ is negligibly small compared with the volume of the equimolar ball; that is, $(\sigma_d/d)r_0^d$, when $L$ is large.

In other words, if    $n = -1 + KL^{d/(d+1)}$ where $K < K_\star$, and $L$ is large, then any droplet in any minimizer must be extremely small. To prove Theorem \ref{thm1b}, it therefore suffices to show that such extremely small drops are impossible in a minimizing order parameter field.  We do this in the next lemma.

\medskip
\begin{lm}\label{uni1}  For all $K >0$, there is a constant $C_K>0$ depending only on $K$ so that if $n \le -1 + KL^{-d/(d+1)}$
and  $m$ is any non uniform minimizer for (\ref{mainmin}), then
$$|\{m > 1- \kappa\}| \ge C_K r_0^d\ .$$
Moreover, $C_K$ is uniformly strictly positive for all $K$ in an interval around $K_\star$.
\end{lm}
\medskip

\medskip

\noindent{\bf Proof:}  We again work with the functional ${\cal G}(w)$ which is defined in (\ref{calGdef})
and related to ${\cal F}(m) - {\cal F}(n)$ by (\ref{FtoG}). Clearly, if  ${\cal F}(m)  < {\cal F}(n)$, then 
the potential $G(w)$, defined in (\ref{Gdef}), must become negative. However, as seen in the proof of Lemma \ref{lb2}, $G(w) < 0$ if and only if $z_- < w < z_+$ where $z_\pm = 2 - 2\delta \pm 2\sqrt{\delta}\sqrt{1-\delta/2}$.

Moreover, $G(w) \ge (n^2 - 1)w^2/2$ for all $w$.  Since $w = m - n$ and by Lemma \ref{lb1}, $m \le 1 + \delta$,
while by definition $n = -1 + \delta$, $w \le 2$. Therefore,
$$G(w(x)) \ge -4\delta(1 - \delta/2)\ $$
for all $x$. 

Define the set $\tilde C$ by $\tilde C = \{ x \ :\ w(x) \ge z_-\ \}$, and define the number $\tilde R$ by 
$$(\sigma_d/d)\tilde R^d  = |\tilde C|\ .$$  We have
\begin{equation}\label{inner}
\int_{\tilde C}\left[{1\over 2}|\nabla w|^2 + G(w)\right]{\rm d}x  \ge 
\int_{\tilde C}G(w){\rm d}x  \ge -4\delta(\sigma_d/d)\tilde R^d\ .
\end{equation}

Define the set $\tilde A$ by $\tilde A = \{ x \ :0 \le \ w(x) \le z_-\ \}$ where the lower bound $0$ is
arbitrary but convenient.  The same argument used to prove Lemma \ref{lb3} shows that

\begin{equation}\label{mid}
\int_{\tilde A}\left[{1\over 2}|\nabla w|^2 + G(w)\right]{\rm d}x    \ge \tilde S\sigma_d\tilde R^{d-1}\ ,
\end{equation}
where $\tilde S = \int_0^{z_-}\sqrt{2G(h)}{\rm d}h$. 

It now follows from (\ref{FtoG}) that ${\cal F}(m) > {\cal F}(n)$ unless
$$4\delta(\sigma_d/d)\tilde R^d \ge   \tilde S\sigma_d\tilde R^{d-1}\ .$$
However,  ${\cal F}(m) > {\cal F}(n)$ is impossible if $m$ is a minimizer. Hence,
$\tilde R \ge \tilde S/(4\delta)$. From (\ref{deldef}), ${\displaystyle
\delta = 2(\sigma_d/d)(r_0^{d+1}/L^d){1\over r_0}}$. 
Hence
$$\tilde R \ge {\tilde S\over 8(\sigma_d/d)}{L^d\over r_0^{d+1}} r_0 \ .$$
The condition $n \le -1 + KL^{-d/(d+1)}$ yields the bound 
$$r_0 \le \left({K\over 2(\sigma_d/d)}\right)^{1/d} L^{d/(d+1)}\ .$$
Thus, there is a constant $C_K$ depending only on $K$ so that  $\tilde R \ge C_K r_0$. 

The final observation to make is $w > z_-$ if and only if $m > 1 - \delta = 2\sqrt{\delta}\sqrt{1 - \delta/2}$.
Therefore, since $\kappa = \delta^{1/3}$, $\tilde C \subset C$ for all $L$ large enough. \qed

\medskip
\noindent{\bf Proof of Theorem \ref{thm1b}}  By Lemma \ref{uni1}, whenever $m$ is a minimizer
for (\ref{mainmin}) with $n < -1 + KL^{-d/(d+1)}$ and $K < K_\star$, the corresponding value of 
$\eta$ is bounded away from zero by a strictly positive quantity depending only on $K$. But by
the remarks preceding  Lemma \ref{uni1}, the $\eta$ value of any minimizer cannot exceed $\bar\eta$,
which tends to zero as $L$ increases.   Hence, for $L$ sufficiently large, there are no non constant
minimizers with $n < -1 + KL^{-d/(d+1)}$ and $K < K_\star$. \qed

\subsection{The proof of Theorem \ref{thm1c}} 

The essential tool here is quantitative version of the isoperimetric inequality. The classical version,
valid in $d=2$, is due to Bonnesen \cite{Bo}, who worked in the setting of convex geometry. 
For the form used here, see \cite{O}. 
To formulate the inequality, let $U$ be a domain in $\R^2$ bounded by  a simply connected
rectifiable curve $|\Gamma|$. Suppose that $\rho_{\rm out}(U)$ is the imfimum of
the radii of  circles
containing $U$, and $\rho_{\rm in}(U)$ is the supremum of the radii of circles contained in $U$. $\rho_{\rm out}(U)$ is called
the {\em outradius} of $U$, and $\rho_{\rm in}(U)$ is called the {\em inradius} of $U$. The inequality of Bonessen
states that
\begin{equation}\label{bon}
|\Gamma|^2 - 4\pi|U| \ge \pi(\rho_{\rm out}(U) - \rho_{\rm in}(U))^2
\end{equation}
where
$|\Gamma|$ denotes ${\cal H}^{1}(\Gamma)$, the $1$ dimensional Haussdorf measure of $\Gamma$. 
That is,
\begin{equation}\label{bo}
{|\Gamma|\over \sqrt{|U|}} \ge \sqrt{\pi}\left(2 + {|\rho_{\rm out}(U) - \rho_{\rm in}(U)|\over \sqrt{|U|}}\right)\ .
\end{equation}

We shall show how this inequality may be used to prove Theorem \ref{thm1c} for $d=2$, and then shall explain how a recent extension by Hall \cite{H}  of Bonnesen's inequality to higher dimensions yields the result for $d>2$. 
The essential points are clearest for $d=2$, and so we begin with that case.

We first have to justify the application of this inequality on the torus. The point is that by properties of the rearrangement employed in \cite{CCELM}, we know the level sets of minimizers $m$ must be connected. If they ``wrap around'' the torus, their perimeter
has a length of at least $2L$, which would give too large a surface contribution to the free energy

To apply  (\ref{bo}), we return to the proof of Lemma \ref{lb3}, and take $U = U_h =\{ m > h\ \}$ for $h_- \le h \le h_+$.
Then, using the notation of section 3,
$$|U_{h_+}| = |C| \qquad{\rm and}\qquad |U_{h_-}| = |C| + |A|\ .$$
The essential point is that in the critical scaling regime, $|A|$ is negligibly small compared to $|C|$
 for large $L$. This is the content of Lemma \ref{lb4}. Therefore, for all $\varepsilon > 0$,
  if $L$ is sufficiently large,
\begin{equation}\label{outer}
 |U_{h_-}| \le (1 + \varepsilon)^2 |U_{h_+}| = (1 + \varepsilon)^2|C|\ .
 \end{equation}
 
 Recall that in section $3$ we have defined $R$ by $\pi R^2 = |C|$.   Now suppose that
\begin{equation}\label{splat} 
\rho_{\rm out}(U_{h_+}) \ge (1+2\varepsilon)R\ .
\end{equation}
Then since for all $h< h_+$, $U_{h_+} \subset U_h$,
\begin{equation}\label{splat2} 
\rho_{\rm out}(U_{h}) \ge (1+2\varepsilon)R\qquad {\rm for\ all} \qquad h_- \le h \le h_+\ .
\end{equation}

On the other hand, since for all $h< h_+$, $U_h \subset U_{h_-}$,  so that
$$|U_h| \le |U_{h_-}| \le \pi(1+\varepsilon)^2R^2\ ,$$
it follows that 
\begin{equation}\label{splat3} 
\rho_{\rm in}(U_{h}) \le (1+\varepsilon)R\qquad {\rm for\ all} \qquad h_- \le h \le h_+\ .
\end{equation}

Combining (\ref{bo}), (\ref{splat2}) and (\ref{splat3}), and letting $\Gamma_h$ denote the boundary of $U_h$,
we have
\begin{equation}\label{splat4} 
{|\Gamma_h|\over \sqrt{|U_h|}} \ge \sqrt{\pi}\left(2 + {\varepsilon\over 1+\varepsilon}\right)
\qquad {\rm for\ all} \qquad h_- \le h \le h_+\ .
\end{equation}

Thus, under the hypothesis (\ref{splat}), 
\begin{equation}\label{splat4} 
|\Gamma_h| \ge 2\pi\left(1 + {\varepsilon\over 2+2\varepsilon}\right)R
\qquad {\rm for\ all} \qquad h_- \le h \le h_+\ .
\end{equation}

Going back to the proof of Lemma \ref{lb3}, one see that effectively, the hypothesis \ref{spalt} would introduce
and extra factor of $(1 + \varepsilon/(2+ 2\varepsilon))$ into the surface tension $S$.  This would increase the
leading order contribution to the free energy over what what obtained in Section 2 with the trial function
$m_{\eta{\rm dr}}$. Therefore, for all sufficiently large $L$, the hypothesis (\ref{spat}) is incompatible with
$m$ being a minimizer. 

The same conclusion can be obtained from Hall's theorem \cite{H} in higher dimension, in essentially the same 
way. The version of Hall's theorem found in Theorem 4.1 of \cite{Ra} is particularly useful for this purpose.
We summarize the discussion in a lemma, using the notation introduced above.

\medskip
\begin{lm}\label{dr1}  Let  $K > K_\star$ and  $n \ge -1 + KL^{-d/(d+1)}$. For any $\varepsilon > 0$
and all $L$ sufficiently large, if
 $m$ is any  minimizer for (\ref{mainmin}), then with $B(x_0,R)$ denoting the ball of radius $R$
  centered on $x_0$, there is a point $x_0$ with
  $${|C \Delta  B(x_0,R)|\over |B(x_0,R)|} \le \varepsilon\ ,$$
  where $\Delta$ denotes the symmetric difference $C\cup B(x_0,R) \backslash (C\cap B(x_0,R))$.
\end{lm}
\medskip

\noindent{\bf Proof of Theorem \ref{thm1c}:}    Let $m$ be any minimizer. After translating, we may
assume that
 $${|C \Delta  B(0,R)|\over |B(0,R)|} \le \varepsilon\ ,$$
 Also, since $K > K_\star$, we know that the value of $R$ must be very close to $\eta_c^{1/d}r_0$.
 Therefore, for all $L$ large enough, we have
 $${|C \Delta  B(0,\eta_c^{1/d}r_0)|\over |B(0,\eta_c^{1/d}r_0)|} \le 2\varepsilon\ .$$
 
 Now, if $ |m(x) - m_{n,\eta}^\sharp(x)\}| > 2\kappa$, it must be that either $ x \in 
 C \Delta  B(0,\eta_c^{1/d}r_0)$, or else $x\in A$.  
 Hence, the set
 $$\{\ x \ :\ |m(x) - m_{n,\eta}^\sharp(x)\}| > 2\kappa\ \}$$
 has a volume no greater than
 $$2\varepsilon   |B(0,r_0)| + |A|\ ,$$
 and we recall that by Lemma \ref{lb4}, $|A|$ is negligibly small compared to $r_0^d$ for all sufficiently large $L$. 
 
 Thus, we obtain
 $$|\{\ x \ :\ |m(x) - m_{n,\eta}^\sharp(x)\}| > 2\kappa\ \}| \le  3\varepsilon   |B(0,r_0)| $$
 for all sufficiently large $L$. Of course we have the globally valid bound
 $|m(x) - m_{n,\eta}^\sharp(x)\}| \le 2$. Therefore,
 $$\int_\Omega |m(x) - m_{n,\eta}^\sharp(x)\}|^4{\rm d}x \le  48\varepsilon   |B(0,r_0)| + 16\kappa^4 L^d\ .
 $$
 Recall that $\kappa = \delta^{1/3}$, and that in the critical scaling regime,
 $$\delta L^d \asymp L^{d/(d+1)} \asymp     |B(0,r_0)| \ .$$
Therefore, for any $\epsilon > 0$
$${1\over |B(0,r_0)|}\int_\Omega |m(x) - m_{n,\eta}^\sharp(x)\}|^4{\rm d}x < \epsilon$$
for all $L$ sufficiently large. \qed


 \def\l{\lambda}
 \def\G{\Gamma}

\section{The construction of good trial functions}\label{trial}

\subsection{A Chapman--Enskog--Hilbert expansion approach} 

As we have seen, the simple trial function $m_{\eta{\rm dr}}$ was sufficient to provide the upper bounds
required here. In fact, it is quite likely that the upper bounds computed in Section 2 are accurate to at least the
first two orders in powers of $L$.   Our goal here is to present a systematic construction of high order trial functions.
This will explain the remarkable efficacy of the simple prescription  $m_{\eta{\rm dr}}$ for (\ref{(1)}), but will
also suggest how one should construct trial functions of similar efficacy for other free energy functionals.
To keep this section concise, we only treat the case $d=2$. This is fully representative, except that the formulas
are much simpler.

For $d=2$, in the critical scaling regime, 
 $r_0 \asymp L^{2/3}$.
 We introduce the scaling parameter
 $\lambda = r_0^{-1}$.
  After the rescaling $x'=\l x$, so that $\O\to \O^\l= \l \O$, so that $|\O^\l|= L^2\lambda^2 =L^{2}r_0^{-2}$, the Euler-Lagrange 
 equation (\ref{el})  becomes
\begin{equation}\label{E.1}
 -\l^2\Delta m +F'(m)+\mu^\l=0\ .
 \end{equation}We will drop the prime on the new coordinate $x$  for sake of simplicity. 
 
 An approximate
solution of order $N$ to the Euler--Lagrange equation is a function $m^{(N)}$ such that
\begin{equation}\label{E.2}
-\l^2\Delta m^{(N)} +f(m^{(N)})+\mu^{(N)}={ O}(\l^{N+1})
  \end{equation}
  Since we are interested in solutions $m$ to the Euler-Lagrange equation such that
$${1\over|\Omega^\l|}\int_{\Omega^\l} m(x) 
  {\rm d}x=n\ ,$$
we require that the constraint on the mass  be approximately
satisfied in the sense that   the approximate solution $m^{(N)}$
satisfies
\begin{equation}\label{acon}  {1\over|\Omega^\l|} \int_{\O^\l}  m^{(N)}(x) {\rm d}x = n + { O}(\l^{N+1})\ .
   \end{equation}
Our aim here is to use an expansion method, based on the Chapman--Hilbert--Enskog expansion of kinetic theory to construct such approximate solution, and to use them as trial function for (\ref{mainmin}), after adding a small constant so that the constraint is exactly satisfied.   

To do this, we first introduce local coordinates in a neighborhood of the curve $\Gamma^{(N)}$,
which will be determined in the course of the expansion.   
Let $s$ denote an arc length parameter along $\G^{(N)}$. The starting point of the parameterization is immaterial.
We denote by $d(x,\G^{(N)})$ the
signed distance of $x$ from $\G^{(N)}$, with  $d(x,\G^{(N)})>0$ when $x$ is in the interior of $\Gamma^{(N)}$
(i.e., the smaller of the two regions into which $\Gamma^{(N)}$ divides the tours).   Define a ``fast'' variable $z$
by $z =  d(x,\G^{(N)})/\lambda$. Then $(s,z)$ give us a system of coordinates on a tubular neighborhood around $\Gamma^{(N)}$.

To construct the approximate solutions, we make the following prescription,
 which has several parts. For any positive integer $N$:

\smallskip
\noindent{(1)} The interfacial curve $\G^{(N)}$ will be a circle of
radius $r^{(N)}$ to be determined at each order, essentially
 by the condition (\ref{acon}). Note that, because of the rescaling the
 radius $r^{(N)}$ is actually measured in units $r_0$ and the
 in the original units the radius of the circle is $r^{(N)}r_0$

\smallskip
\noindent{(2)} The chemical potential $\mu^{(N)}$ has an expansion of the form
$$\mu^{(N)} = \l\mu_1 + \l^2\mu_2 + \cdots +\l^N\mu_N\ .$$

\smallskip
\noindent{(3)} We then construct
\begin{equation}\label{chen}
m^{(N)}=\bar m\left({d(x,\Gamma^{(N)})\over
\lambda}\right)+\sum_{n=1}^N\l^n\left[h_n+\phi_n\right] \
\end{equation}
   where:{ \it (i)}  $m_0$ is the approximation to $\bar m$ introduced in Section 2,
{\it (ii)} $\phi_j$ will be a bounded continuous function that is nearly constant away from $\Gamma^{(N)}$.
{\it (iii)}
$h_j$ is a function that has the form
   $$h_j\left({d(x,\G^{(N)})\over \l}\right) \ ,$$
where 
   on the right, $h_j(z)$ denotes a rapidly decaying function of the variable $z$
   (The notation is such that the symbol $h_n$ plays two roles, but this should cause no confusion.)

   By using the local coordinates $(s,z)$ around the curve $\Gamma^{(N)}$, 
we write the Laplacian as
\begin{equation}\label{laplcur}
   \l^2\Delta f={\partial^2\over \partial z^2}f+{\l K^{(N)}\over(1-z\l K^{(N)})}{\partial\over \partial z}f
+ {\l^2\over(1-z\l K^{(N)})^2}{\partial^2\over \partial s^2}\
\end{equation}
and
\begin{equation}\label{curv}
    {1\over 1-z\l K^{(N)}}=\sum_{n=0}(-1)^n{(\l K^{(N)}z)^n\over n}:=\sum_{n=0}\l^n k_n
\ .
   \end{equation}
    Note that our simplifying assumption that $\G^{(N)}$
is a circle amount to assume that $K^{(N)}$ does not depend on $s$,
thus dropping a few terms of higher order in $\l$ in (\ref{laplcur}).
Note also, for future reference that the area element in these
coordinates is
\begin{equation}\label{areaelt}
    (1 + K^{(N)} \l z) \l{\rm d}z{\rm d}s\ 
    \end{equation}
and recall that the  surface tension, $S$,
is given by
\begin{equation}\label{virial}{S}=\int_{\Bbb R}  |\bar m'(z)|^2{\rm d}z =2
\int_{\Bbb R}  F(\bar m'(z)){\rm d}z \end{equation}
    
We expand $f=F'$ as
\begin{eqnarray}    
    f(m^{(N)})&=&f(\bar m)+ f'(\bar m)\sum_{n=1}^N\l^n(h_n+
\phi_n)\nonumber\\
    &+&{1\over 2}f''(\bar
m)\left(\sum_{n=1}^N\l^n(h_n+\phi_n)\right)^2+{1\over 3!}f'''(\bar
m)\left(\sum_{n=1}^N \l^n(h_n+\phi_n)\right)^3\ .\nonumber\\
    \end{eqnarray}

Our goal is to show that  we can choose the $r^{(j)}$, $\mu_j$,
$h_j$ and $\phi_j$ so that $m^{(N)}(r) $ is a high order approximate
solution of (\ref{E.1}).  At the $j$-th stage, to determine $\phi_j$,
we will solve an equation  in $L^\infty(\R)$, and to determine
$h_j$, we will solve an equation  in $L^2(\R)$. The Fredholm
criterion for solvability of this equation will relate $\mu_j$ to
$r^{(j)}$, and the constraint equation (\ref{acon}) will then
determine $r^{(j)}$.

\subsection{The first order approximate solution} \label{ex2}

To see how this goes, we insert
$$m^{(1)}(r) =   \bar m\left({d(x,\G^{(1)})\over \l}\right) + \l(h_1+\phi_1)$$
into (\ref{E.1}), collect terms by order in $\l$, solve the resulting equations to find
$r_1$, $\mu_1$, $h_1$ and $\phi_1$.

 The result is:
 $$-\bar m'' + F'(\bar m) + \l\left(-K^{(1)}\bar m'  - h_1''  - \phi_1'' + F''(\bar m)h_1 +
 F''(\bar m)(\phi_1 + \mu_1\right) + {\cal O}(\l^2)\ .$$
Since, by the definition of $\bar m$ the term multiplied by $\l^0$
vanishes, we need to equate to $0$ the coefficient of the term of
order $\l$. This corresponds to put
 \begin{equation}\label{fo1}
   -K^{(1)}\bar m'  + (- h_1'' + F''(\bar m)h_1) +
 (-\phi_1'' + F''(\bar m)\phi_1) + \mu_1 = 0\ .
    \end{equation}

 Define the operator ${\cal L}$ by
 $${\cal L} = -{{\rm d}^2\over {\rm d}z^2} + F''(\bar m)\ .$$
Also define ${\cal L}_0$ by
 $${\cal L}_0 = -{{\rm d}^2\over {\rm d}z^2} + F''(1) = -{{\rm d}^2\over {\rm d}z^2}  + {1\over \chi}\ .$$
In the present example, $1/\chi =2$, and so ${\cal L}^{-1}$ is given by convolution  with the Helmholtz Green's function
$$R(z) = {1\over 2\sqrt{2}}e^{-\sqrt{2}|z|}\ .$$ ${\cal L}_0^{-1}$ maps constants into constants and
preserve the parity properties of functions.

 Then we can rewrite (\ref{fo1}) as
  \begin{equation}\label{fo2}{\cal L}h_1 = K^{(1)}\bar m'  + (F''(1) - F''(\bar m))\phi_1 - {\cal L}_0\phi_1 - \mu_1\ .
     \end{equation}
The first two terms on the right hand side are rapidly decaying in $z$. In order for the entire right hand
side to rapidly decay, and thus belong to $L^2(\R)$,  we require that the last to terms cancel. That is, we require
$${\cal L}_0\phi_1 + \mu_1 = 0\ . $$
This is solved by
 \begin{equation}\label{chiandmu}
     \phi_1 =-\chi \mu_1\ ,
     \end{equation}
     and so (\ref{fo2}) reduces to
 \begin{equation}\label{fo2b}{\cal L}h_1 = K^{(1)}\bar m'  + (1 - \chi F''(\bar m))\mu_1\ .
     \end{equation}

  We have
\begin{equation}\label{0.1}\phi_1= {\cal L}_0^{-1}\mu_1=-{1\over\chi}\mu_1
\end{equation}
The equation for $h_1$ becomes
\begin{equation}\label{0.1.2} {\cal L}h_1=K^{(1)}\bar m'+(1-\chi f'(\bar m)\mu_1=0
\end{equation}

The null space of ${\cal L}$ is spanned by $\bar m'$, and so
 the Fredholm criterion says that (\ref{0.1.2}) is solvable if and only if
 $K^{(1)}\bar m'  + (1 - \chi f'(\bar m))\mu_1$
 is orthogonal to $\bar m'$.
     Multiply by $\bar m'$ and integrate. Using the fact that $\int_\R f'(\bar m)\bar m'{\rm d}z = 0$, and $ \int_\R \bar m'{\rm d}z = -2$, we obtain,
 $$K^{(1)}\left(\int_\R(\bar m')^2{\rm d}z\right) = -2\mu_1\ .$$
     
We see that to leading order in $\lambda$, the curvature must be constant, and so $\Gamma^{(N)}$ must be a circle,
since we are considering values of $n$ that are close to $-1$.   Let $r^{(1)}$ be the radius of the circle, which is, as yet, undetermined.

Using (\ref{virial}) to express the integral in terms of $S$,  we obtain
\begin{equation}\label{mu1form}
      \mu_1 = -{K^{(1)}S\over 2}\ .
      \end{equation}

Moreover, since $\bar m(z) = -{\rm tanh}(z/\sqrt{2})$, it is easy to see that both
$\bar m'(z)$ and $(1 - \chi f'(\bar m))$ are proportional to ${\rm sech}^2(z/\sqrt{2})$, and so with
$\mu_1$ given by (\ref{mu1form}), the right hand side of (\ref{0.1.2}) vanishes identically, and
we see that $h_1 = 0$.   From (\ref{0.1}) and (\ref{mu1form}), we have
\begin{equation}\label{phi1form}
      \phi_1 = \chi{K^{(1)}S\over 2}\ .
      \end{equation}
 Finally, we determine $r^{(1)}$ using the approximate constraint (\ref{acon}). Since $K^{(1)} = 1/r^{(1)}$, we have that
$$m^{(1)} =
\bar m\left({d(x,\G^{(1)})\over \l}\right) +  {\chi S\over 2r^{(1)} } \ .$$

Toward this end, note that by (\ref{areaelt}),
$${1\over |\O^\l|}\int_{\O^\l}  \bar m\left({d(x,\G^{(1)})\over \l}\right) {\rm d}x =
-1+ 2\pi {(r^{(1)})^2\over |\O^\l|} + { O}(\l^2)\ .$$ Thus,
\begin{eqnarray}\label{phi}
{1\over |\O^\l|}\int_{\O ^\l} m^{(1)}(x){\rm d}x&=& -1 + 2\pi
{(r^{(1)})^2\over|\O^\l|} +   \l\phi_1+ {O}(\l^2)\nonumber\\
     &=& -1 + 2\pi
{(r^{(1)})^2\over|\O^\l|} +   \l {\chi S\over 2r^{(1)} } +
{O}(\l^2)\nonumber\\
      &=&  n +   { O}(\l^2)=-1+ {2\pi \over |\O^\l|}+O(\l^2)\
.\nonumber\\
      \end{eqnarray} This yields
\begin{equation}\label{r1form}
     2\pi [(r^{(1)})^2-1] +   \l|\O^\l|{\chi S\over 2r^{(1)}}=0\ ,
     \end{equation}
which is a cubic equation determining $r^{(1)}$.   One could solve it explicitly -- it is, after all, a depressed cubic.
 However, the important point is that the of the three roots, only two are positive, and the largest one is
 exactly the radius determined by the Biskup--Kotecky--Chayes prescription.
 
 To see this, we compute the free energy  ${\cal F}(m^{(1)})$.  First note that, with the
scaling we used, we have
$${\cal F}(m^{(1)}) = \l^{-2}\int_{\O^\l}  [{\l^2\over 2} |\nabla m^{(1)}|^2
+F(m^{(1)})]{\rm d}x\ .$$ 
       By using the expression of $m^{(1)}$ and (\ref{areaelt}) to
pass to the variables $(s,z)$, we get
$${\cal F}(m^{(1)}) = \l^{-1}2\pi S r^{(1)}+ {\l^{-2}|\O^\l|\over 2
\chi}\l^2\phi_1^2.$$
From (\ref{phi}) and (\ref{equimolar}) we compute
$$\phi_1= {2\pi(1-(r^{(1)})^2)\over \l|\O^\l|}+O(\l).$$
Therefore
$$
{\cal F}(m^{(1)}) =\l^{-1}\left(2\pi r^{(1)}S +{\l^{-3}L^{-2}\over
\chi}2\pi^2(1-(r^{(1)})^2)^2\right) + {O}(1)
$$
This is exactly the phenomenological free energy of \cite{BCK} for the Free energy functional (\ref{(1)}).
Also, the Euler--Lagrange equation for it reduces to (\ref{r1form}).   Hence $m^{(1)}$, 
with $r^{(1)}$ chosen to be the minimizing solution of  (\ref{r1form}) is essentially exactly the trial function
we used to obtain the upper bound in Section 2. The only difference is a slight adjustment of the additive constant so that the constraint in (\ref{mainmin}) is exactly satisfied.

The crucial feature in the free energy function (\ref{(1)}) that is responsible for this is that in this case we found $h_1 = 0$.

 \subsection{The second order approximate solution} \label{ex2}

Going on to second order is not difficult since $h_1 =0$, and $\phi_1$ is an explicit constant.
The next order displays some new features, but once the calculations are carried out to second order, it will be clear how to go on to arbitrary order.

The equations for $h_2$ and $\phi_2$ are
$$-{d^2\over dz^2}(h_2+\phi_2)-K^{(2)}\bar m'+f'(\bar m)(h_2+\phi_2) {1\over 2}f''(\bar m)(h_1+\phi_1)^2+ \mu_2 = 0$$
As before, write
$${\cal L}\phi_2 =  {\cal L}_0 \phi_2  + (f'(\bar m) - f'(1))\phi_2\ ,$$
and we then have
$${\cal L}h_2  = -\left[ K^{(2)}\bar m' +
(F''(\bar m) - f'(1))\phi_2\right]  -{\cal L}_0 \phi_2  -
{1\over 2}f''(\bar m)\phi_1^2 - \mu_2\ .$$
The terms in square brackets  are rapidly decaying as long as $\phi$ is bounded. To eliminate the other terms, we choose
$${\cal L}_0 \phi_2 =  -{1\over 2}F'''(\bar m)\phi_1^2 - \mu_2\ .$$
This is easily solved in $L^\infty(\R)$ using the Helmholtz Green's function. With this choice of $\phi_2$,
we determine $h_2$ through
$${\cal L}h_2  = -K^{(2)}\bar m' -
(F''(\bar m) - F''(1))\phi_2\ .$$
The right hand side must be orthogonal to $\bar m'$. Recall that $(f'(\bar m) - f'(1))$ is even, and in fact is
a multiple of $\bar m'$. Since $f''(\bar m) = 6\bar m$, which is odd, only the even part of $\phi_2$
is relevant in the solvability condition. But clearly, the even part of $\phi_2$ is given by
$$(\phi_2)_{\rm even} = -\chi\mu_2\ .$$
Thus, the solvability condition becomes
$$K^{(2)}\left(\int_\R(\bar m'(z))^2{\rm d}z\right)^2 + {1\over \chi}\chi 2\mu_2 = 0\ .$$

As before, with this choice,
$$K^{(2)}\bar m' +
(f'(\bar m) - f'(1))(\phi_2)_{\rm even}  =0\ ,$$
since it is a multiple of $\bar m'$, and the equation for $h_2$ reduces to
$${\cal L}h_2 = -(f'(\bar m) - f'(1))(\phi_2)_{\rm odd}\ .$$
The mass condition is
\begin{eqnarray}
{1\over |\O^\l|}\int_\O  m^{(2)}(x){\rm d}x &=& -1 + 2\pi
{(r^{(2)})^2\over |\O^\l|} +   \l{\chi S\over 2r^{(1)}
}+\l^2(\phi_2)_{\rm even} + { O}(\l^3)\nonumber\\
      &=&  n  +   { O}(\l^3)=-1+
{2\pi \over |\O^l|}+O(\l^3)\ .\nonumber\\
      \end{eqnarray}
       By  choosing
$r^{(2)}=r^{(1)}+\l r_2$, recalling that $(\phi_2)_{\rm even}={\chi
S\over 2 r^{(2)}}$ we get
$$r_2=-{\chi S\over 8(r^{(1)})^2}.$$
The even part of $\phi^{(2)}$ is the most significant term among the new correction. However, it is        
a constant, so that even keeping the most significant term at second order, we still have a trial function
 of the simple type considered in Section 2.  For this reason, one can expect that the upper bounds obtained in section 2 are in fact sharp not only in the leading order, but in the first two orders in powers of $L$.

The procedure can be continued along the same lines to higher order
thus producing approximate solutions to the Euler-Lagrange equations.   However, what is probably more significant is that it can be applied to other free energy functional with non local interactions for which it
       will not be the case that $h_1 =0$, and hence the construction of a suitable trial function is not so simple.

\end{document}